\documentstyle[sprocl]{article}

\def\Journal#1#2#3#4{{#1} {\bf #2}, #3 (#4)}

\def\NPB{{\em Nucl. Phys.} B}

\def\PRL{\em Phys. Rev. Lett.}
\def\PRD{{\em Phys. Rev.} D}

\def\JMP{\em J. Math. Phys.}
\def\be{\begin{equation}}
\def\ee{\end{equation}}
\def\bea{\begin{eqnarray}}
\def\eea{\end{eqnarray}}
\begin{document}

\title{QUANTIZING CANONICAL GRAVITY IN THE REAL DOMAIN}

\author{R. LOLL}

\address{INFN Firenze, Largo E. Fermi 2,\\
50125 Firenze, Italy}
 
\maketitle\abstracts{We advocate an alternative description of
canonical gravity in 3+1 dimensions, obtained by using as the
basic variable a {\it real} variant of the usual Ashtekar connection
variables on the spatial three-manifold. With this ansatz, no
non-trivial reality conditions have to be solved, and the Hamiltonian
constraint, though non-polynomial, can be quantized rigorously in a
lattice regularization.}

\section{A fresh look at the classical structure}

There are several possibilities of formulating the classical
Hamiltonian theory of pure Einstein gravity. The traditional one,
proposed by Arnowitt-Deser-Misner, is in terms of a canonical pair
$(g_{ab},\pi^{ab})$ of a Riemannian three-metric and its conjugate
momentum. Introducing local, rotational SO(3)-degrees of
freedom, one obtains a closely related formulation, based on a variable
pair $(E^a_i,K_a^i)$, where the (inverse, densitized) three-metric is
expressible as a function of the triad $E^a_i$, $g^{ab}=E^a_i E^{bi}$,
and its conjugate momentum is the extrinsic curvature of the
three-manifold $\Sigma$, with one of its spatial indices converted to
an internal one.

Starting from this latter formulation, one may perform a canonical
transformation and end up with yet another canonical pair of
variables $(A_a^i,E^a_i)$, where $A_a^i$ is now an SO(3)-valued
{\it connection} variable. To be precise, we will be interested
in two variants of this approach. In the first case, the gene\-rator of
the canonical transformation is $i\int \Gamma_a^i E^a_i$, and one
obtains the so-called Ashtekar variables \cite{aa} $(A^{Ash},E)$, with
$A^{Ash}=\Gamma +iK$, whereas in the second case one uses the generator
$\int \Gamma_a^i E^a_i$, and obtains a real version $(A^{Re},E)$
of the Ashtekar variables, with $A^{Re}= \Gamma +K$. (The connection
variables $\Gamma_a^i=\Gamma_a^i (E)$ are defined by the vanishing
of the covariant derivatives, ${\cal D}_a(\Gamma) E^b_i=0$.) These may
be understood as two special cases of the canonical transformation
\cite{fb} $E^a_i=E^a_i$,  $A_a^i=\Gamma_a^i +\beta K_a^i$, for a
non-vanishing constant $\beta$. For the corres\-ponding Poisson
brackets, one derives $\{ A_a^i(x),E^b_j(y)\}=\beta\
\delta^i_j\delta_a^b \delta^3(x,y)$.

In this new formulation, there are seven first-class constraints,
\begin{equation}
\begin{array}{rcl}
G_i & := &  \nabla_a E^a_i\equiv\partial_a E^a_i +\epsilon_{ijk}
   A_a^j E^{ak} = 0, \\
V_a & := &  F_{ab}^i E^b_i\equiv 
  (\partial_a A_b^i -\partial_b A_a^i +\epsilon^{ijk} A_{aj}
   A_{bk} )\ E^b_i= 0, \\
H   & := & \epsilon^{ijk} E^a_i E^b_j ( F_{abk}(A) -
  (\frac{1}{\beta^2}+1)\ R_{abk} (\Gamma) )=0,
\end{array}\label{one}
\end{equation}
\noindent the Gauss law constraints $G_i$ (familiar from
Yang-Mills theory), the three spatial diffeomorphism constraints $V_a$
and the Hamiltonian constraint $H$. As indicated, $F$ denotes the
curvature of the connection $A$, and analogously, $R$ the curvature
of the connection $\Gamma$. Since $\Gamma_a^i$ is a rather complicated
and non-polynomial function of the triads $E^a_i$, its
appearance in (\ref{one}) spells trouble for the quantum
theory. One may circumvent this by choosing (\`a la
Ashtekar) $\beta=\pm i$, resulting in a 
polynomial form for $H$ and thus all the constraints. 

An unusual feature of this choice is that it makes
the fundamental variable $A_a^i$ complex. In order to eliminate
the unphysical modes this introduces, one has to impose a sufficient
number of conditions on the (a priori arbitrary) SO(3,C)-valued
connections $A_a^i$. One way of doing this is to
require that $A_a^i +A_a^{i*}=2 \Gamma_a^i(E)$. Conditions of this
type are often called ``reality conditions". This sounds simple, but
one should keep in mind that they define a highly non-linear
subspace of the original space of gauge connections and triads.

One may of course keep the classical formulation real, by setting
$\beta =\pm 1$, say. This does not affect the functional
form of the Gauss law and spatial diffeomorphism constraints, but the
Hamiltonian retains a non-polynomial piece. Thus it seems as 
if one would run into difficulties similar to those occurring  
in the canonical quantization based on the metric variable pair
$(g_{ab}, \pi^{ab})$, where the complicated structure of the
Hamiltonian has basically led to an impasse.
However, some particular features of the connection representation
allow one to take a somewhat different stance, as I will explain in
the next section.  

\section{Quantization}
  
In the quantization of Ashtekar gravity, one roughly speaking
proceeds as follows. The local SO(3)-gauge symmetry is eliminated by
working with explicitly gauge-invariant wave functions $\Psi(A)$, 
taking the form of Wilson loop functio\-nals $\Psi(\gamma)={\rm Tr}\
P\exp\int_\gamma A$, where $\gamma$ denotes a closed curve in
$\Sigma$. Note that the information about $A$ is encoded in such
objects in a rather singular way. Since the wave functions are
labelled by the ``extended objects" $\gamma$, they carry a natural
action of the diffeomorphism group (by moving the loop argument),
which may be used to construct diffeomorphism-invariant wave
functions. 

In the usual Ashtekar approach, the remaining Hamiltonian constraint 
is given by $H=\epsilon^{ijk} E^a_i E^b_j F_{abk}$. Following Dirac, it
is translated into the quantum theory as an operator condition
$\hat H\Psi=0$ on physical states $\Psi$ (subject to operator-ordering
problems and an appropriate regularization and renormali\-zation). 
One robust feature of $\hat H$ is the existence of a large
class of solutions to $\hat H\Psi=0$ in representations where $\hat
A$ acts by multiplication. Solutions of this type were found 
already several years ago \cite{js,rs}, and are labelled by smooth,
non-intersecting loops $\gamma$. However, there are two
problems associated with them: firstly, they are obtained modulo
imposition of quantum analogues of the reality conditions
(which have not even been formulated in the loop representation);
secondly, there are indications that they lie in a physically trivial
sector of ``states without volume" (see below). A reasonably large
class of solutions overcoming both of these difficulties has not yet
been found.

One is therefore led to reconsider the real connection approach,
correspon\-ding to the choice $\beta=\pm 1$. As a first step, one
brings the non-polynomial terms in the Hamiltonian $H=\epsilon^{ijk}
E^a_i E^b_j F_{abk}- H^{\rm pot}$ into the form of
polynomials modulo powers of the determinant of the metric
\cite{rl} (recall $\det g=|\det E|$), 
\be
H^{\rm pot}=  \frac{1}{(\det E)^2}\eta_{a c d}\eta_{e g h}
(E^{c}_k E^{d}_l E^{g}_m E^{h}_n
- 2 E^{c}_m E^{d}_n E^{g}_k E^{h}_l) 
E^{b}_k E^{f}_m 
(\nabla_{b} E^{a}_l)(\nabla_{f} E^{e}_n),
\label{two}
\ee
where the determinant is a cubic function of the triads,
\be
\det E= \frac{1}{6}\eta_{abc}\epsilon^{ijk} E^a_i E^b_j E^c_k.
\label{three}
\ee
What enables us to quantize (\ref{two}) in spite of its
non-polynomiality is the fact that in the loop representation the
classical volume function ${\cal V}=\int d^3x\sqrt{ |\det E|}$ can be
quantized self-adjointly, and a basis of eigenstates of
$\hat{\cal V}$ can be constructed in terms of linear
combinations of Wilson loop states. (Note that this has no
analogue in the metric representation.) This is true both in the
continuum \cite{rs2} and in a lattice-regularized version of the theory
\cite{rl2}. One consequence of this is that -- at least on the lattice
-- any function of $\det E$ can be quantized exactly by going to a
basis of eigenstates and defining it in terms of its eigenvalues. In
the case of inverse powers of $\det E$, one of course has to take care
that zero-eigenstates do not occur.

At this point the reader should be warned that some crucial
differences exist between the fixed-lattice approach where the
lattice discretizes space itself, and the continuum formalism where
the quantum states are labelled by all possible {\it embedded} loops
\cite{rs,almmt}.
In the first case, the spatial diffeomorphism symmetry is destroyed
by the discretization, and one still has to take a continuum limit
(in which the diffeomorphism symmetry may be restored).
By contrast, in the continuum the full diffeomorphism
group of $\Sigma$ is still present, and apparently no additional 
continuum limit is necessary, because one is considering all possible
embedded lattices (supporting loop states) simultaneously. 
At an intermediate stage, the properties of operators acting on
loop states can be similar in both approaches (e.g. the
spectrum of $\hat{\cal V}$ is always discrete),
but the details of calculations and interpretation tend to be different.

Let us go back to the discussion of the Hamiltonian. Taking into
account the anti-symmetrizations in the numerator of 
$H^{\rm pot}$, one may suspect that the entire expression (\ref{two})
is not actually as divergent as the $(\det E)^{-2}$-factor suggests.
This expectation is in fact correct: consider
the continuum expression
\be
\frac{1}{2\sqrt{\det E}}\eta_{abc} \epsilon^{ijk} E_j^b E_k^c
:=e_a^i.
\label{four}
\ee
Up to a density factor, the $e_{ai}$ are the inverses of the
$E^a_i$, $E^a_i e_{aj}=\sqrt{\det E}\delta_{ij}$. On the other
hand, $e_a^i$ may be expressed as the functional derivative of
the total volume with respect to $E_i^a$ \cite{tt}, 
$e_a^i(x)=2\{A_a^i(x),\int d^3x \sqrt{\det E}\}$, which
is non-singular even for $\det E=0$.
A similar construction can be performed on the lattice, i.e. one
can define a lattice link operator that in the continuum limit
to lowest order in the lattice spacing $a$ reduces to $e_{ai}$.
Since $H^{\rm pot}$ contains a fourfold product
of the $e_a^i$, this implies that $H$ can in fact be made well-defined
on a large number (or possibly all) of the lattice states. 

Given this discretization of the real connection approach, one can
now try to solve $\hat H\Psi=0$, without having to 
implement any reality conditions. (Note that one is really trying to
find solutions to this set of equations {\it to lowest order in
$a$}). To what extent
this approach is computationally feasible, and what further
approximations may be necessary is currently being explored. 

We conclude by remarking that the difference between the metric and
connection approaches to quantum gravity emerging from the above
discussion is not that in the latter 
one can altogether avoid the appearance of non-polynomial quantities,
but rather that one can bring non-polynomiality into a form where it can be 
controlled.

\end{document}